# Recombination in polymer-fullerene bulk heterojunction solar cells

Sarah R. Cowan[§], Anshuman Roy[§] and Alan J. Heeger[*]

*Center for Polymers and Organic Solids, University of California, Santa Barbara, California, 93106, USA*

**Abstract**

Recombination of photogenerated charge carriers in polymer bulk heterojunction (BHJ) solar cells reduces the short circuit current ($J_{sc}$) and the fill factor (*FF*). Identifying the mechanism of recombination is, therefore, fundamentally important for increasing the power conversion efficiency. Light intensity and temperature dependent current-voltage measurements on polymer BHJ cells made from a variety of different semiconducting polymers and fullerenes show that the recombination kinetics are voltage dependent and evolve from first order recombination at short circuit to bimolecular recombination at open circuit as a result of increasing the voltage-dependent charge carrier density in the cell. The "missing 0.3V" inferred from comparison of the band gaps of the bulk heterojunction materials and the measured open circuit voltage at room temperature results from the temperature dependence of the quasi-Fermi-levels in the polymer and fullerene domains – a conclusion based upon the fundamental statistics of Fermions.

[§] These authors contributed equally to this work.
[*] Corresponding author: ajhe@physics.ucsb.edu

## I. INTRODUCTION

The power conversion efficiency (*PCE*) of a solar cell is given by the well-known relation $PCE = J_{sc}V_{oc}FF/P_{in}$ where $J_{sc}$ is the short circuit current, $V_{oc}$ is the open circuit voltage, *FF* is the Fill Factor and $P_{in}$ is the incident solar power. Since recombination results in loss of photogenerated charge carriers, acquiring an understanding of the mechanisms governing recombination is critical for increasing $J_{sc}$ and *FF* and thereby increasing the solar cell performance. Using a detailed balance approach, Shockley and Queisser[1] showed that the open circuit voltage of a solar cell is maximum when the photogenerated charges recombine only radiatively. For bulk heterojunction (BHJ) solar cells made from blends of semiconducting polymers and fullerenes, the recombination mechanisms are mostly non-radiative.[2] Thus, overcoming such recombination can, in addition, increase the open circuit voltage.

The recombination mechanisms in polymer BHJ solar cells are, however, far from clear. For P3HT:PC$_{60}$BM[3] cells, contradictory explanations based on both first order (monomolecular)[4-7] and bimolecular[8-10] recombination have been proposed, but have met with only limited success in explaining the current-voltage characteristics. Recently, for PCDTBT:PC$_{71}$BM solar cells with power conversion efficiency greater than 6%,[11] Shockley-Read-Hall recombination at interfacial traps[7] was proposed as the dominant mechanism.

Monomolecular and bimolecular recombination are terms which require precise definition. Here, we use monomolecular recombination synonymously with any first order process. The intensity dependent current-voltage studies described here determine process order alone. Monomolecular recombination historically refers to either Shockley-Read-Hall (SRH) recombination or geminate recombination. Transient photoconductivity measurements[12] carried out on operating solar cells established that geminate recombination is not the dominant mechanism in P3HT:PC$_{60}$BM and PCDTBT:PC$_{71}$BM solar cells.

Shockley-Read-Hall recombination is a first order recombination process in which one electron and one hole recombine through a trap state or recombination center. Impurities in the fullerene and polymer materials and incomplete phase separation (interfacial defects that function as traps) are likely to contribute to a trap-based recombination. The fundamental assumption which makes SRH recombination first order is the time delay between the capture of the first charge and the second charge. The quick initial capture of electrons (density, $n_e$) creates a reservoir of stationary trapped electron charge with which mobile holes, $n_h$, can recombine. The trap based mechanism transforms the recombination from a bimolecular process with incident light intensity, $I$, $R_{BI} \propto n_e(I) \cdot n_h(I)$, to a first order process, $R_{SRH} \propto n_{e,trap} \cdot n_h(I)$.

In addition to SRH recombination, first order recombination can originate, for example, from charge carrier concentration gradients set up through the depth of the bulk heterojunction solar cell.[13] Although constant charge generation throughout the bulk heterojunction material at steady state reduces the concentration gradient, a large imbalance can exist near the electrodes, e.g. a higher density of electrons. Such an imbalance will enable mobile holes, $n_h$, to recombine in the presence of a significant excess of electrons, $n_{e,exc}$, via a first order process: $R_{1stBI} \propto n_{e,exc} \cdot n_h(I)$.

Here we ask the following questions: Is recombination in polymer solar cells monomolecular, bimolecular, or a combination of the two? What is the effect of applied voltage on the recombination kinetics? We arrive at answers to these important questions by measuring the current-voltage characteristics over a range of different illumination intensities and temperatures using polymer BHJ solar cells made from three different semiconducting polymers (PCDTBT, P3HT, KP[3,14]) and two different fullerenes ($PC_{60}BM$ and $PC_{71}BM$[3]). Our results reveal that the kinetics of recombination for polymer BHJ solar cells depend on the external voltage applied to the device: The current density versus voltage (*J-V*) curves are limited by first order recombination from the short circuit condition to the maximum power point and evolve to bimolecular recombination in the range of voltages from the maximum power point to the open circuit condition. Furthermore, we find a universal dependence of the open circuit voltage ($V_{oc}$) for polymer BHJ solar cells on incident light intensity; $\delta V_{oc} = (k_B T/e) \ln(I)$, where *I* is the incident light intensity, $k_B$ is the Boltzmann constant, *T* is the absolute temperature and *e* is the electron charge. The slope of $\delta V_{oc}$ vs. ln(*I*) demonstrates that bimolecular recombination dominates for applied voltages near $V_{oc}$. For comparison, we also probe the incident light intensity dependence and temperature dependence of the current-voltage characteristics of a p-i-n junction hydrogenated amorphous silicon (a-Si) solar cell, where the recombination has been ascribed to trap sites at the p/i and i/n interfaces.[15]

To probe the kinetics of recombination, solar cells were fabricated with the following compositions: PCDTBT:$PC_{71}BM$ (1:4), P3HT:$PC_{60}BM$ (1:0.7), and KP:$PC_{60}BM$ (1:3); details are provided in Appendix A. In addition, measurements were carried out on commercial amorphous silicon (a-Si) solar cells obtained from Contrel Technology Co. Ltd. (Taiwan) and a single crystal silicon (c-Si) solar cell NREL certified for light source calibration from PV Measurements, Inc. *J–V* characteristics were collected while illuminating the solar cells over a range of intensities from 0.4 mW/cm² to 100 mW/cm². The spectrum of the incident light was adjusted for every value of intensity to closely mimic the AM 1.5G spectrum.

## II. INTENSITY DEPENDENCE OF THE CURRENT-VOLTAGE CURVES

Fig. 1(a) shows the current-voltage characteristics of the PCDTBT:PC$_{71}$BM solar cell for incident light intensities ranging from 0.4 to 100 mW/cm$^2$. The total current density flowing through the solar cell is a function of the incident light intensity ($I$) and the applied voltage ($V$), and is given by the sum of the dark current ($J_{dark}$) and photogenerated current ($J_{photo}$):

$$J_{photo}(I,V) = J(I,V) - J_{dark}(V) \qquad (1)$$

The photocurrent in Eq. 1 can be written as $J_{photo}(I,V) = edG(I)P_C(I,V)$, where $G(I)$ is the photon flux absorbed by the solar cell per unit volume, $d$ is the distance between the electrodes and $P_C(I,V)$ is the charge collection probability. As is evident from Fig. 1(a), the current becomes independent of the applied voltage around -0.5 volts (reverse bias). Hence, assuming a reverse saturation current such that $G(I) = J_{photo}(V = -0.5$ volts$)$:

$$P_C(I,V) = \left| \frac{J_{photo}(I,V)}{J_{photo}(I,-0.5V)} \right| \qquad (2)$$

We note that for PCDTBT: PC$_{71}$BM, the internal quantum efficiency approaches 100% so that $P_C$ is known to approach unity at short circuit.[11]

Fig. 1(b) shows the charge collection probability as a function of applied voltage. The data from all the curves shown in Fig. 1(a) collapse onto a universal voltage dependence in the range of applied voltages from -0.5V to approximately 0.7V; i.e. close to the maximum power point. Hence, $P_C(I, V_{applied}) \approx P_C(V)$, independent of intensity from short circuit to $V_{applied} > V_{MPP}$ where $V_{MPP}$ is the voltage at the maximum power point.

Given the linear variation of $J(-0.5V)$ with incident light intensity shown in the inset of Fig. 1(b) and the collapsed collection probability curve in the range of voltages from -0.5 to 0.7 volts, we conclude that the photocurrent in this voltage range is linearly dependent on intensity. Therefore in this regime, the recombination is dominated by a first order (monomolecular) mechanism.

We note that because the devices reported here exhibit relatively high efficiency, most of the charge generated at short circuit, up to 90%, is swept out of the device prior to recombination and collected as current in the external circuit. Thus, measurements at short circuit, where the internal field is high and the charge carriers are efficiently swept out, are not ideal for the study of recombination. In reverse bias, nearly 100% of the photogenerated carries are swept out and recombination plays an insignificant role. Therefore, our conclusion of first order recombination intimately relies on the collapse of all the data (see Fig. 1(b)) from reverse bias to the maximum power point. Even at the maximum power point, the probability of recombination remains independent of intensity.

The data at short circuit show that $J \propto I^\alpha$, where $\alpha = 1$ to high accuracy, as emphasized in Fig. 5 in Appendix B, where the data are plotted on a log-log scale and fit to a power law. The interpretation of the exponent $\alpha$ varies in literature. One can find numerous attempts to account for $\alpha$ in terms of monomolecular versus bimolecular recombination,[16-18] and there are a number of reasons given in the literature why $\alpha$ may be less than 1. Powers less than 1 could result from bimolecular recombination,[19] space charge effects,[19] variations in mobility between the two carriers[20] or variations in the continuous distribution in the density of states.[20] We emphasize, however, that $\alpha \equiv 1$ when all carriers are swept out prior to recombination.

Plots similar to Fig. 1(a) for the intensity dependent current-voltage characteristics of solar cells made from P3HT:PC$_{60}$BM, KP:PC$_{60}$BM, and the a-Si (p-i-n junction) are shown in Fig. 6 of Appendix B.

Fig. 2(a) shows the linear dependence of the short circuit current of the P3HT:PC$_{60}$BM, KP:PC$_{60}$BM, and a-Si (p-i-n junction) solar cells on incident light intensity. The collection probabilities for P3HT:PC$_{60}$BM, KP:PC$_{60}$BM, and a-Si (p-i-n junction) solar cells are plotted as a function of voltage and intensity in Figs. 2(b)-(d). Again, the collection probability for incident light intensities that vary over an order of magnitude collapse in the voltage range from -0.5 volts to near the maximum power point. This collapse again indicates that throughout this voltage range, the photocurrent increases linearly with intensity and implies intensity-independent recombination. Hence, for all these BHJ solar cells, first order (monomolecular) recombination dominates for the range of applied voltages from -0.5 volts to near the maximum power point.

Beyond the maximum power point, however, the charge collection probability becomes dependent on the incident light intensity; see Fig. 1(b) and Figs. 2(b)-(d). The spread in the collection probability curves for various incident light intensities is most evident at the open circuit voltage, the externally applied voltage at which the total current is zero. As we show below, this variation with light intensity arises from a change in the recombination kinetics with voltage, evolving from first order (monomolecular) recombination for voltages up to the maximum power point to bimolecular recombination when the external current is zero, i.e. the open circuit condition; see Fig. 1(b).

## III. INTENSITY DEPENDENCE OF THE OPEN CIRCUIT VOLTAGE

The light intensity dependence of $V_{oc}$ provides independent and complementary information on the details of the recombination processes from that obtained from $J_{sc}$ (and *FF*). Under open circuit conditions, the current is zero; all photogenerated carriers recombine within the cell. Thus, recombination studies near open circuit are particularly sensitive to the details of the recombination mechanism.

Fig. 3(a) shows that the open circuit voltage varies logarithmically (ln($I$)) with light intensity and that all the curves of $\delta V_{oc}$ vs ln($I$) for polymer:fullerene BHJ solar cells have the same slope, ($k_B T/e$). For the a-Si solar cell, $\delta V_{oc} \sim 1.7(k_B T/e)$ ln($I$)). An NREL certified silicon calibration cell has slope $2(k_B T/e)$. Fig. 7 in the Supplementary Information shows the variation of $V_{bi}$, the voltage at which the photocurrent is zero, with incident light intensity. Within the margins of experimental error, the dependence of $V_{bi}$ on light intensity is identical to that of the open circuit voltage; $\delta V_{bi} = (k_B T/e)\ln(I)$.

## IV. ANALYSIS OF THE CROSS-OVER FROM MONOMOLECULAR TO BIMOLECULAR RECOMBINATION

The internal voltage within the device, given by the difference $V_{bi} - V$, drives the carriers to the electrodes and determines the timescale for the sweep-out of carriers, $\tau_s = d^2 / 2\mu(V_{bi} - V)$, where $\mu$ is the charge carrier mobility, $d$ is the distance between the electrodes, $V$ is the applied voltage, and $V_{bi}$ is the built-in potential (see Appendix C for details of the derivation). At a given voltage, competition between sweep out and recombination determines the carrier density available for recombination within the device.

The increased carrier density with decreasing internal voltage (decreasing carrier sweep-out) causes the transition from monomolecular to bimolecular recombination kinetics. At short circuit, the photocurrent, $J_{sc} = J_{photo} = edGP_C(V=0)$. In the optimized solar cell limit, the collection probability $P_C \to 1$ at short circuit, and $J_{sc}$ is dominated by drift current from the photogenerated carriers in the internal field:

$$J_{drift} = J_{photo} = 2en_{sc}\mu V_{bi}/d = edG \tag{3a}$$

where $n_{sc}$ is the electron (or hole) density at short circuit. This equation can be rewritten in terms of the timescale for the sweep-out of carriers, $\tau_s$:

$$G = n_{sc}/\tau_s \tag{3b}$$

The recombination rate ($R$) at open circuit voltage can be written as a sum of two terms (see Appendix D for details of the derivation):

$$R(V_{oc}) = G = \frac{n_{oc}}{\tau_r} + \gamma n_{oc}^2 \tag{4}$$

where $\tau_r$ is the monomolecular recombination lifetime, $n_{oc}$ is the electron (or hole) density within the device at open circuit, and $\gamma$ is the bimolecular recombination coefficient. The ratio of the two terms summed in Eq. (4) will determine the carrier density, $n_e$, or alternatively, the magnitude of the bimolecular recombination coefficient, at which the recombination mechanism transitions from monomolecular to bimolecular kinetics; this cross-over is given by $\gamma n_e > 1/\tau_r$.

The balance of charge carriers within the device at any voltage is determined by the continuity equation ($e = 1.6 \times 10^{-19}$ Coulombs):

$$\frac{1}{e}\frac{\partial J}{\partial x} = G - R \tag{5}$$

At open circuit, $dJ/dx = 0$ and Eq. (5) reduces to $G = R(V_{oc})$. From Eqs. (3b) and (4), we obtain:

$$\frac{n_{sc}}{\tau_s} = n_{oc}\left(\frac{1}{\tau_r} + \gamma n_{oc}\right) \tag{6}$$

Since bimolecular recombination dominates at open circuit, $\gamma n_{oc} > 1/\tau_r$, implying that $n_{oc}/n_{sc} > \tau_r/\tau_s$. The ratio $\tau_r/\tau_s > 10$, as obtained from transient photoconductivity measurements on operating solar cells.[12]

Short circuit current, $J_{sc}$, estimates the carrier density in the device under steady state conditions (AM 1.5G solar spectrum); see Eq. (3a). For PCDTBT:PC$_{71}$BM solar cells,[11] $J_{sc} = 11$ mA/cm$^2$. Thus, $n_{sc} \approx 10^{15}$ cm$^{-3}$ (assuming $\mu = 10^{-3}$ cm$^2$/V-s).[21-22] Using $\tau_r \approx 10^{-6}$ s obtained directly from transient photoconductivity measurements carried out on operating solar cells[12] and assuming that $\gamma n_{oc}/n_{sc} \gg 1/\tau_r$, we find $\gamma \leq 10^{-12}$ cm$^3$/s. The inferred value for $\gamma$ is significantly smaller than the magnitude obtained from the Langevin expression,[23-25] $\gamma = e\mu/\varepsilon$.

Many measurement methods have been utilized to study recombination, including steady state photocurrent,[8] integral mode time of flight,[24] transient photovoltage (TPV), transient absorption spectroscopy,[25] and photo-charge extraction with linearly increasing voltage[26] measurements. Street[7] has noted that near short circuit, bimolecular recombination is suppressed relative to monomolecular (trap-induced) recombination by the low density of carriers. In the BHJ nanostructure, the polymer and fullerene domains have dimensions of approximately $2 \times 10^{-6}$ cm, and the Coulomb escape radius is also of this order. Thus, the domain size is of order $10^{-17}$ cm$^{-3}$. Bimolecular recombination is suppressed near short circuit because the number of electrons or holes per domain is < 1 (and often zero). Recent papers on recombination at open circuit have attempted to explain the orders of magnitude discrepancy between experimental measurements and the Langevin model. Szmytkowski[27] proposed that an effective medium approximation of the dielectric permittivity of the bulk heterojunction blend may reduce the effective recombination coefficient by orders of magnitude, and Groves and Greenham[9] conclude from Monte Carlo simulations that the effects of energetic disorder, domain sizes, and the electron-hole mobility mismatch are not enough to describe the reduction, and that deep carrier trapping may explain the magnitude of the recombination rate. In contradiction to steady state experiments, a charge density dependent bimolecular recombination rate is necessary to fit experimental TPV and charge extraction data obtained from the P3HT:PCBM donor-acceptor system.[28] Deibel, Wagenpfahl, and Dyakonov[13] offer an explanation of the apparent contradiction in the literature between the charge density dependent recombination rate in transient experiments and a charge density independent recombination rate in steady state experiments due to charge carrier concentration gradients.

## V. EFFECT OF RECOMBINATION ON THE OPEN CIRCUIT VOLTAGE

The recombination mechanism governs the extent to which the incident light intensity modulates the open circuit voltage. When a polymer solar cell is under illumination at open circuit, the applied voltage equals the difference between the quasi-Fermi-levels within the polymer and fullerene phase separated domains. From this observation, we obtain the following expression for the open circuit voltage:

$$V_{oc} = \frac{1}{e}\left(E_{LUMO}^{Fullerene} - E_{HOMO}^{Polymer} - \Delta\right) - \frac{kT}{e}\ln\left(\frac{n_e n_h}{N_c^2}\right) \quad (7)$$

where $n_e$ and $n_h$ are the electron and hole densities in the fullerene and polymer domains at open circuit, and $N_c$ is the density of conduction states at the band edge of the polymer and fullerene, assumed here for the purpose of argument to be equal. The energy shift, $\Delta$, in the first term of Eq. (7) originates from disorder within the solution cast and phase separated polymer and fullerene regions as sketched in the inset to Fig. 3(b).[29]

The commonly accepted value, $V_{oc} \approx E_{LUMO}^{Acceptor} - E_{HOMO}^{Donor}$, is obtained from Eq. (7) only at $T = 0$ K. The validity of the first term in Eq. (7) has been verified for a number of polymer:fullerene BHJ systems, but with a reduction of 0.3 V of previously unknown origin.[30] At finite $T$, because of the fundamental statistics of Fermions, the quasi-Fermi levels move away from $E_{LUMO}^{Acceptor}$ and $E_{HOMO}^{Donor}$, respectively, and into the gap above the polymer HOMO energy level and below the fullerene LUMO energy level. The resulting reduction in $V_{oc}$ is given by the second term in Eq. (7) and is the origin of the "missing 0.3 V."

In the limit where bimolecular recombination is dominant ($\gamma > 1/(n_{oc}\tau_r)$), $n_e n_h = (n_{oc})^2 = G/\gamma$. When substituted into Eq. (7), this results in $\delta V_{oc} = (k_B T/e)\ln(I)$ + constant, where $I$ is the incident light intensity. In Fig. 3(a), we plot the light intensity dependence of the open circuit voltage for all the polymer:fullerene BHJ solar cells described in Fig. 1 and Fig. 2. The data demonstrate that the slope is of $\delta V_{oc}$ vs. ln(I) is equal to $k_B T/e$ within the measurement error. In addition, we include in Fig. 3(a) the light intensity dependence of $V_{oc}$ as reported by others obtained using different semiconducting polymers in the BHJ material.[31-33] We find that for all these different polymer-fullerene systems, the slope is equal to $k_BT/e$. This universality highlights the generality of bimolecular recombination kinetics at open circuit in polymer BHJ solar cells.

If first order (monomolecular) recombination were the dominant mechanism over the full range of applied voltages from short circuit to open circuit, the "collapsed" J-V curve would look qualitatively the same as that shown in Fig. 1(b). However, for monomolecular recombination $n_e$ and $n_h$ (at open circuit) would each be proportional to the intensity, and the slope of $\delta V_{oc}$ vs ln(I) would be $2(k_BT/e)$. In Fig. 3(a), the slope for the amorphous silicon solar cell is $1.7(k_BT/e)$, suggesting that recombination at open circuit is a combination of monomolecular and bimolecular processes. As also shown in Fig. 3(a), the slope for the crystalline silicon solar cell is $2(k_BT/e)$, implying that monomolecular (Shockley-Read-Hall) recombination is dominant even at open circuit. Thus, the slope of $\delta V_{oc}$ vs. ln(I) provides a straightforward method for distinguishing monomolecular and bimolecular recombination.

Using the bimolecular limit of Eq. (7), with $n_e n_h$ proportional to intensity, the data in Fig. 3(a) can be collapsed onto the universal curve shown in Fig. 3(b). All the polymer BHJ cells show identical intensity dependence, $\delta V_{oc} = (k_BT/e)\ln(I)$. To obtain this universal curve, we calculated the magnitude of carrier generation rate for each of our solar cells using measured absorption coefficient spectra gathered from ellipsometry. Because the precise value for the disorder induced shift, $\Delta$, is expected to be different for each polymer, there is uncertainty in the precise values for $\gamma(N_c)^2$; for example, a shift of $\Delta = 0.1$ eV along the ordinate of Fig. 3(b) corresponds to a decrease in $N_c$ by an order of magnitude.

Nevertheless, ignoring the $\Delta$-shift and using $\gamma \approx 10^{-12} \, cm^3/s$, reasonable values for $N_c$ are obtained for the different polymers ($\sim 10^{19}$-$10^{20} \, cm^{-3}$).

The temperature dependence of $V_{oc}$ for PCDTBT:PC$_{71}$BM solar cells is plotted in Fig. 4. Fig. 4(a) shows $V_{oc}$ vs. light intensity at different temperatures (the temperature of the cell was controlled during measurement using a Peltier cooler/heater). The lines overlaid on the data are not fits to the data, but are lines predicted by Eq. (7) from the measured temperatures.

In Fig. 4(b), we replot the data from Fig. 4(a) to show the linear dependence of $V_{oc}$ with temperature at various light intensities. The dashed lines, predicted by Eq. (7), fit well to the data and predict an interfacial band offset $V_{oc} = \frac{1}{e}\left(E_{LUMO}^{Fullerene} - E_{HOMO}^{Polymer} - \Delta\right) = 1.25V$. Best linear fits to the $V_{oc}$ vs. temperature data at varying intensities (not shown in Fig. 4(b)) give $V_{oc}(T = 0K) = 1.27V \pm 0.02V$. This value is in agreement with cyclic voltammetry measurements of the HOMO energy of PCDTBT[34] (-5.5 eV) and of the LUMO energy of PCBM[35] (-4.3 eV). The difference provides a value for the interfacial band gap. Fig. 4(b) provides an independent measure of the electronic structure obtained in situ, with an accuracy of 0.02 eV. Similar measurements and analysis for 4 other polymer systems verify the validity of this analytic method as shown in Fig. 8 in Appendix B.

The temperature-dependent measurements indicate that for these BHJ solar cells, under the incident light intensities in this experiment, the quasi-Fermi levels are *not* pinned at the interfacial band gap. The voltage difference between the interfacial band gap and the $V_{oc}$ measured at room temperature results from thermal shifts in the quasi-Fermi-levels (see Eq. (7)). Fig. 4(b) suggests that it would be possible to measure the full interfacial gap at low temperatures. However, in practice, reduced mobility at low temperatures will localize the photogenerated carriers and result in a non-linear reduction in the current and the open-circuit voltage.

For the past 5 years, the reduction of $V_{oc}$ (the "missing 0.3V" described by Brabec *et al* [30]) compared to the value estimated by the interfacial gap has remained a mystery. In Fig. 4(b) and Fig. 8, we demonstrate that this loss is the result of the temperature dependence of the quasi-Fermi-levels in the polymer and fullerene domains – a conclusion based upon the fundamental statistics of Fermions.

## VI. SUMMARY AND CONCLUSIONS

To summarize, experiments on polymer BHJ solar cells made from a variety of different materials reveal that voltage-dependent charge carrier recombination evolves from being first order (monomolecular) in carrier density at short circuit to being second order (bimolecular) at open circuit. Interfacial trap states between the polymer and fullerene domains likely determine the first order (monomolecular) recombination that is dominant from the short circuit condition to the maximum power point. For the polymer BHJ solar cells, the densities of electrons and holes at 100 mW/cm$^2$ (AM1.5G solar spectrum) and the magnitude of the bimolecular recombination coefficient lead to the crossover from monomolecular recombination at short circuit to bimolecular recombination at open circuit. We expect that for higher trap densities, the Fill Factor would decrease and the recombination kinetics would remain monomolecular over the full range of applied voltages (even at open circuit).

The temperature/intensity dependence of $V_{oc}$ shown in Fig. 4(b) demonstrates that the intrinsic open circuit voltage is reduced significantly from the commonly accepted value by temperature dependent shifts in the energies of the quasi-Fermi-levels.

Reducing the trap density through control of the phase separated morphology and the composition of the interface is a major opportunity for the science of BHJ materials. By reducing the interfacial trap density, it should be possible to increase the charge extraction (sweep-out) efficiency, enabling an increase in the thickness of the photoactive layer and thereby further increasing the short circuit current without negatively affecting the Fill Factor, all steps toward polymer BHJ solar cells with power conversion efficiency beyond 10%.

**ACKNOWLEDGMENTS**

This research was supported by the US Army General Technical Services (LLC/GTS-S-09-1-196), the Air Force Office of Scientific Research (AFOSR FA9550-08-1-0) and the Department of Energy (DOE ER46535). We thank Dr. D. Waller (Konarka Technologies) for providing the materials. We thank Dr. Robert Street (Palo Alto Research Center) for transmitting to us, through important discussions, his fundamental insight into the operation of BHJ solar cells. We thank Dr. Thuc-Quyen Nguyen for stimulating comments.

**APPENDIX A: Solar Cell Fabrication and Intensity Measurements**

Polymer-fullerene solar cells were fabricated using blends of four polymers and two fullerene derivatives, [6,6]-phenyl C61 butyric acid methyl ester ($PC_{60}BM$) and [6,6]-phenyl

C71 butyric acid methyl ester (PC$_{71}$BM). Devices using poly[3-hexylthiophene] (P3HT), the copolymer poly[N-9"-hepta-decanyl-2,7-carbazole-alt-5,5-(4',7'-di-2-thienyl-2',1',3'-benzothiadiazole) (PCDTBT), and the Konarka polymer, KP. Devices of reproducible quality were fabricated on indium tin oxide (ITO) coated substrate with the following structure: ITO-coated glass substrate / poly(3,4 ethylenedioxythiophene) : poly(styrenesulfonate) (PEDOT:PSS) / polymer:fullerene blend / TiOx / Al.

P3HT:PC$_{60}$BM films were cast from a solution of P3HT:PC$_{60}$BM (1:0.7) in chlorobenzene solvent with a polymer concentration of 10 mg/mL and 3 volume percent of the additive (1,8 diiodooctane). BHJ films were annealed after casting at 70C for 10 minutes to drive out excess solvent. Devices were not annealed after deposition of the aluminum electrode. PCDTBT:PC$_{71}$BM blend films were cast from a solution of PCDTBT:PC$_{71}$BM (1:4) in a 1,2 dichlorobenzene : chlorobenzene solvent mixture (3:1) with a polymer concentration of 7 mg/mL. BHJ films were annealed after casting at 60C for 1 hour to drive out excess solvent. Devices were not annealed after deposition of the aluminum electrode. KP:PC$_{60}$BM films were cast from a solution of KP:PC$_{60}$BM (1:3) in a dichlorobenzene : chlorobenzene solvent mixture (3:1) with a polymer concentration of 10 mg/mL. BHJ films were annealed after casting at 60C for 1 hour to drive out excess solvent. Devices were not annealed after deposition of the aluminum electrode. An amorphous solution processable TiOx layer was cast from solution onto all the devices as a buffer layer and an optical spacer.[36] The TiOx film was annealed for 10 minutes at 80C in air to oxidize the film and evaporate solvent. A 100 nm aluminum electrode was vacuum deposited. Devices were encapsulated for testing in air.

**Table 1:** Initial device efficiencies for devices used in this study

| Material | $J_{sc}$ (mA/cm$^2$) | $V_{oc}$ (V) | FF | PCE (%) |
|---|---|---|---|---|
| P3HT:PC$_{60}$BM | 7.3 | 0.64 | 0.58 | 2.70 |
| KP:PC$_{60}$BM | 9.2 | 0.63 | 0.58 | 3.33 |
| PCDTBT Mw = 100 kDa:PC$_{71}$BM | 10.6 | 0.88 | 0.68 | 6.28 |
| PCDTBT Mw = 58 kDa:PC$_{71}$BM | 10.7 | 0.86 | 0.61 | 5.55 |

Current density–voltage (*J–V*) characteristics of the devices were measured using a Keithley 236 Source Measure Unit. Solar cell performance used a Newport Air Mass 1.5 Global (AM 1.5G) full spectrum solar simulator with an irradiation intensity of 100 mW/cm$^2$. The 100 mW/cm$^2$ spectrum of incident light was spectrum and intensity matched with an Ocean Optics USB4000 spectrometer calibrated for absolute intensity via a deuterium tungsten halogen calibration standard lamp with NIST-traceable calibration from 350-1000 nm. Initial device performance is listed in Table 1. Over the 2 month period of device testing, the devices degraded a maximum of 10% from their initial efficiency. The intensity of the lamp was modulated with a series of 2 neutral density filter (NDF) wheels of 6 filters apiece, allowing for 35 steps in intensity from 100 mW/cm$^2$ – 0.4 mW/cm$^2$. Intensity of light transmitted through the filter was independently measured via a power meter. Error is introduced while modulating the full solar spectrum with "grey" filters, which non-linearly reduce the solar spectrum, especially at high filter optical densities. Scatter in the data specific to the density filters result in error in the fit of $V_{oc}$ vs ln(*I*) of ±0.001 – equivalent to an uncertainty of ±12 degrees in temperature or ± 0.04 in the "slope." Therefore, external temperature measurement is necessary to reduce experimental error in the fitting – as implemented below.

Solar illumination increases the temperature of the device while under illumination. We carefully measure the device temperature after allowing the device to come to equilibrium temperature under full illumination. The area of a large metal heat sink (optical table) is illuminated by the solar simulator (illumination area = 232 cm$^2$). By allowing the solar simulator to heat this large volume heat sink to an equilibrium temperature, adjusting the intensity incident on the device via neutral density filters becomes a relatively small perturbation on the equilibrium temperature. The light intensity on the solar cell (substrate area = 2.25 cm$^2$, active area = 0.15 cm$^2$) is varied by inserting a neutral density filter (area = 3.1 cm$^2$) during *J-V* testing. Thus, the total power absorbed by the heat sink does not change by the temporary insertion of the neutral density filter. The device is thermally anchored directly to the metal heat sink, and the actual temperature is measured in situ by a thermocouple. Equilibrium temperature of ~35°C is reached after approximately 30 minutes under the solar simulator under 1 sun illumination. The dark current is determined at approximately the same temperature by measurement immediately after device testing, while the cell is still hot. In situ thermocouple measurements indicate the temperature difference to be at most a few degrees (2-3°C) between the 1 sun illuminated measurement and the dark current measurement.

**APPENDIX B: Supplemental Data**

Fig. 5 shows the data from Fig. 2(a) plotted on a log-log scale. The data are fitted with a power law. The linear least squares errors to the fit are given in the Inset. The BHJ data at short circuit show that $J \propto I^{\alpha}$, where $\alpha = 1$ to high accuracy. Fig. 6(a)-(d) shows the intensity dependent current-voltage characteristics of (a) the PCDTBT Mw = 100kDa:$PC_{71}BM$ solar cell, (b) the P3HT:$PC_{60}BM$ solar cell, (c) the KP:$PC_{60}BM$ solar cell, and (d) the a-Si p-i-n junction solar cell for incident light intensities ranging from 0.4 to 100 mW/cm$^2$.

The intensity dependence of $V_{bi}$, the voltage where the photocurrent = 0, is plotted in Fig. 7. The photocurrent was calculated by subtracting the diode dark current from the measured device current. $V_{bi}$ was found by linearly interpolating the photocurrent data near the x axis. Data are the colored circles, and modeled lines are overlaid. Modeled lines account for the offset from $V_{oc}$ and use the same slope as those obtained from the $V_{oc}$ fits: $(k_BT/e)$ for the polymers, 1.7 $(k_BT/e)$ for a-Si, and 2 $(k_BT/e)$ for single crystal Si.

The temperature dependence of $V_{oc}$ for several polymer:fullerene solar cells is plotted in Fig. 8(a)-(d). Fig. shows the linear dependence of $V_{oc}$ with temperature at various light intensities. The dashed lines, predicted by Eq. (7), fit well to the data. Data in Fig. 8(c-d) on the PPV materials from L.J.A. Koster, V.D. Mihailetchi, R. Ramaker and P.W.M. Blom, Appl. Phys. Lett. **86**, 123509 (2005), L.J.A. Koster, Ph.D. Thesis, University of Groningen, 2007, and V. Dyakonov, Appl. Phys. A-Mater. Sci. Process. **79**, 21-25 (2004)..

**APPENDIX C: Derivation of the Sweep-Out Time**

Carrier mobility is defined by the relationship between the drift velocity of carriers in an electric field by the formula:

$$v_d = \mu E \tag{AC.1}$$

We relate the drift velocity to the distance charge travels in the electric field, $x$, and the characteristic sweep-out time, $\tau_s$, by the relation:

$$v_d = \frac{x}{\tau_s} \tag{AC.2}$$

We assume a uniform electric field, and that the average collection length is half that of the cell thickness assuming charge is generated uniformly throughout the sample. A modification of this simple model could incorporate both the charge generation profile and dispersive transport. Therefore, within the model

$$x = \frac{d}{2} \tag{AC.3}$$

where $d$ is the thickness of the bulk heterojunction layer. Under the same uniform electric field approximation, $E = V_{int}/d$. The internal potential due to band bending ($V_{bi}$) and the external electric field ($V$), is $V_{int} = V_{bi} - V$. Therefore, Eq. (AC.3) can be solved for the characteristic sweepout time:

$$\tau_s = \frac{d^2}{2\mu V_{int}} \tag{AC.4}$$

**APPENDIX D: Derivation of the Recombination Rate Due to Shockley-Read-Hall Recombination and Langevin Recombination**

At open circuit, $J = dJ/dx = 0$ -- i.e., the net current and the current flux are both equal to zero. Hence, at steady state, we can eliminate the current flux as well as time dependent terms from the electron and hole continuity equations, and consider both interfacial recombination mediated by traps along with bimolecular recombination at steady state to obtain:

$$\frac{\partial n_e}{\partial t} = G - (R_{e,trap} - G_{e,trap}) - R_b = 0 \qquad (AD.1)$$

$$\frac{\partial n_h}{\partial t} = G - (R_{h,trap} - G_{h,trap}) - R_b = 0 \qquad (AD.2)$$

$$\frac{\partial n_{e,trap}}{\partial t} = (R_{e,trap} - G_{e,trap}) - (R_{h,trap} - G_{h,trap}) = 0 \qquad (AD.3)$$

where, $n_e$ is the density of electrons in the LUMO of fullerene and $n_h$ is the density of holes in the HOMO of polymer, $G = G_e = G_h$ is the generation rate of excited electron-hole pairs due to the absorption of incident photons, $R_b$ is the rate of bimolecular recombination, $R_{e,trap}$ is the rate at which electrons fall into (interfacial) traps from the LUMO of fullerene, $G_{e,trap}$ is the rate at which trapped electrons are thermally ejected form a trap into the LUMO of fullerene, $R_{h,trap}$ is the rate at which holes are lost to traps, and $G_{h,trap}$ is the rate at which trapped holes are ejected from traps into the HOMO of polymer. We assume that all excited electron-hole pairs diffuse to the polymer-fullerene interface and dissociate into electrons in the LUMO of fullerene and holes in the HOMO of polymer. This is a reasonable assumption for at least some polymer solar cells (PCDTBT:PC$_{71}$BM, P3HT:PC$_{60}$BM) where the IQE at short circuit has been found to approach 100%.[10]

Additional assumptions:

1.  The traps are located primarily at the interface of polymer and fullerene domains.

2. Electrons fall into interfacial traps from the LUMO of fullerene. Similarly, holes from the HOMO of polymer recombine with electrons in occupied interfacial traps.

3. Traps are neutral when unoccupied and negatively charged when occupied by electrons -- important for the sign convention used in the above equations.

4. An electron moving at its thermal velocity, $v_e$, and hovering inside the capture cross-section of a trap, $\sigma_e$, will get trapped.

5. Occupied traps do not capture a second electron -- higher order terms would be required to describe this process, which we consider unnecessary at this point. Moreover, the electron-electron Coulomb repulsion would inhibit double occupancy.

$$R_{e,trap} = \sigma_e v_e \left( n_{trap} - n_{e,trap} \right) n_e \quad (AD.4)$$

$$R_{h,trap} = \sigma_h v_h n_{e,trap} n_h \quad (AD.5)$$

$$R_b = \gamma \left( n_e n_h - n_i^2 \right) \quad (AD.6)$$

where $n_i^2 = n_e^0 n_h^0$ and $n_i$ is the intrinsic carrier density at equilibrium, $n_e^0$ is the density of electrons in the LUMO of fullerene and $n_h^0$ is the density of holes in the HOMO of polymer at equilibrium.

We can write the following rate expressions for the ejection rates of trapped electrons and holes into the LUMO(fullerene) and HOMO(polymer) as:

$$G_{e,trap} = \beta_e n_{e,trap} \quad (AD.7)$$

$$G_{h,trap} = \beta_h \left( n_{trap} - n_{e,trap} \right) \quad (AD.8)$$

The coefficients $\beta_e$ and $\beta_h$ determine the extent of thermal carrier ejection from traps. We assume that the magnitudes of these coefficients are independent of applied voltage and remain unchanged with illumination. To determine $\beta_e$ and $\beta_h$, we go to Eq. (AC.1) and (AC.2) and set G = 0 (i.e., dark state when $n_e = n_e^0$ and $n_h = n_h^0$) under steady state and at equilibrium. We note that the bimolecular recombination term drops out and we obtain

$$n_{e,trap} = n_{trap} \frac{1}{\frac{\beta_e}{\sigma_e v_e n_e^0}+1} = n_{trap}\frac{1}{\exp\left(\frac{E_{trap}-E_F}{kT}\right)+1} \quad (AD.9)$$

The above equation assumes that at equilibrium, the density of trapped electrons is given by the Fermi distribution contained in the second equality above. Also, at equilibrium the density of electrons in the LUMO of fullerene is given by:

$$n_e^0 = N_c \exp\left(-\frac{E_{LUMO}-E_{trap}}{kT}\right) \quad (AD.10)$$

Substituting Eq. (AC.9) into Eq. (AC.10), we can obtain expressions for $\beta_e$ and $\beta_h$. We can then use Eq. (AC.3) to obtain an expression for the density of electrons in traps at steady state, $n_{e,trap}$; upon plugging this expression into the steady state form of either Eq. (AC.1) or (AC.2), we obtain

$$R = G = \left(n_e n_h - n_i^2\right)\left[\gamma + \frac{1}{\tau_e\left(n_e + n_{e,trap}\right)+\tau_h\left(n_h + n_{h,trap}\right)}\right] \quad (AD.11)$$

where,

$$\begin{aligned}n_e^t &= N_v \exp\left[-\left(E_{trap}-E_{HOMO}\right)/kT\right] \\ n_h^t &= N_c \exp\left[-\left(E_{LUMO}-E_{trap}\right)/kT\right]\end{aligned} \quad (AD.12)$$

The thermal charge densities and the thermal population of traps can be assumed to be small, such that $n_i \ll n_e(n_h)$, and $n_{e,trap}(n_{h,trap}) \ll n_e(n_h)$. Thus at open circuit, defining $\tau_r = \tau_e + \tau_h$,

$$R(V_{oc}) = G(V_{oc}) = \frac{n_{oc}}{\tau_r} + \gamma n_{oc}^2$$

(AD.13)

**Figure Captions**

**Figure 1: (a)** Current-voltage characteristics of PCDTBT:PC$_{71}$BM solar cells as a function of incident light intensity; **(b)** Charge collection probability: Photocurrents measured for the various intensities in (a) have been normalized with the photocurrent at -0.5 volts. The two ovals highlight voltage ranges where monomolecular and bimolecular recombination kinetics are dominant. Inset: The magnitude of current density at -0.5 volts plotted against incident light intensity.

**Figure 2: (a)** Short circuit current plotted against incident light intensity for different solar cells; Charge collection probability of solar cells made from **(b)** P3HT:PC$_{60}$BM, **(c)** KP:PC$_{60}$BM, **(d)** p-i-n junction amorphous silicon.

**Figure 3: (a)** Open circuit voltage as a function of incident light intensity; OC$_1$C$_{10}$:PPV : DPM-10 (purple),[33] PCDTBT Mw = 100 kDa : PC$_{71}$BM (royal blue), PCDTBT Mw = 58 kDa : PC$_{71}$BM (green), BEH:PPV : PC$_{60}$BM (tan),[32] MDMO:PPV : PC$_{60}$BM (yellow),[31] P3HT : PC$_{60}$BM (black), KP : PCBM (dark blue), amorphous silicon (red), single crystal silicon (pink). **(b)** Universal curve showing $\delta V_{oc}$ as given by Eq. (7). Inset: Schematic diagram of the density of states in the band "tails" and the intensity dependent quasi-Fermi energies (at $T=0$ K) as the "tails" are filled by photoexcited electrons (in the fullerene component) and holes (in the polymer component). At finite temperatures, the quasi-Fermi energies move into the gap (see Eq. (7)).

**Figure 4: (a)** Logarithmic dependence of $V_{oc}$ with incident light intensity with slope $k_B T/e$, cell temperature modulated. **(b)** Linear dependence of $V_{oc}$ with temperature in a PCDTBT:PC$_{71}$BM solar cell.

**Figure 5:** Short circuit current plotted against incident light intensity for different solar cells (Fig. 2(a)) on a logarithmic scale. Data is fit to a power law, with fit powers and linear least squares error inset.

**Figure 6:** Intensity dependent current-voltage characteristics for solar cells: **(a)** PCDTBT Mw = 100kDa:PC$_{71}$BM, **(b)** P3HT:PC$_{60}$BM, **(c)** KP:PC$_{60}$BM, **(d)** a-Si p-i-n junction.

**Figure 7:** Intensity dependence of $V_{bi}$, the voltage where the photocurrent = 0.

**Figure 8:** Temperature and light intensity dependent $V_{oc}$ measurements estimate the interfacial band gap for (a) P3HT:PC$_{60}$BM, (b) KP:PC$_{60}$BM, (c) MDMO:PPV:PC$_{60}$BM,[31] (d) BEH:PPV:PC$_{60}$BM.[32]

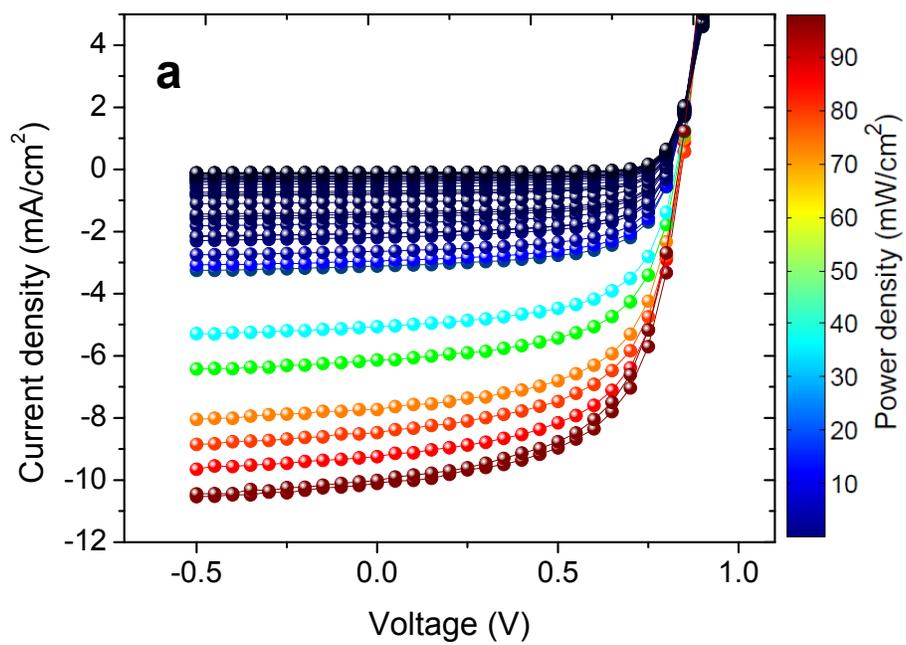

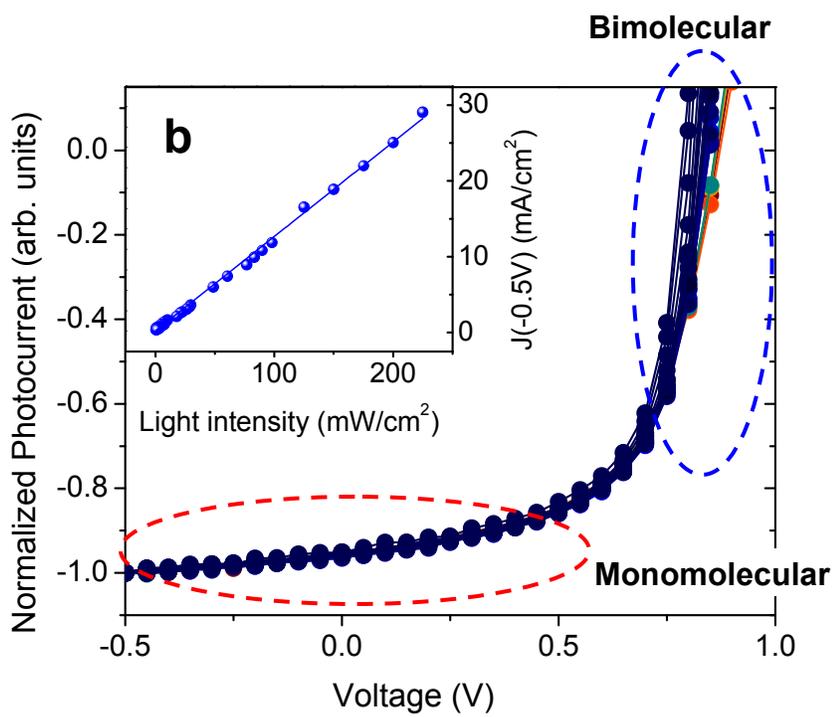

**Figure 1**

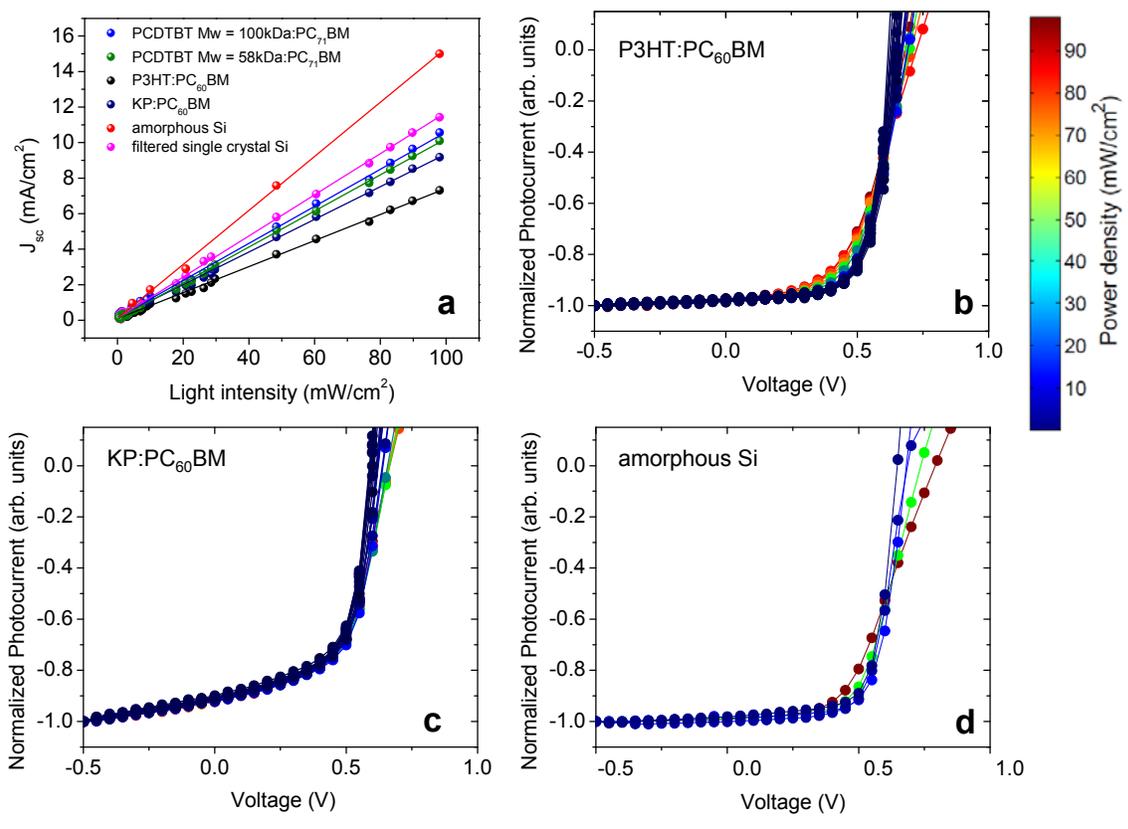

**Figure 2**

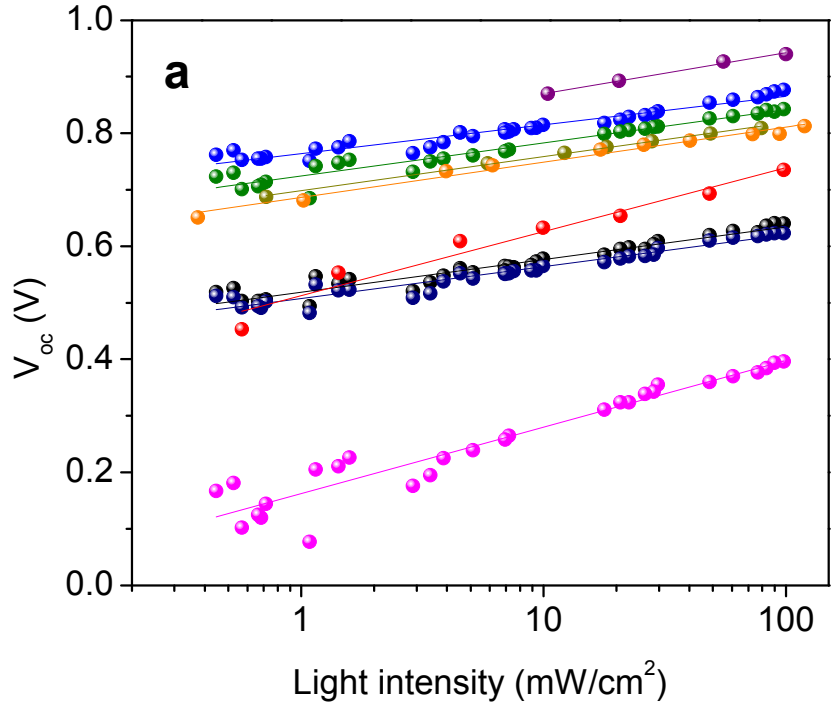

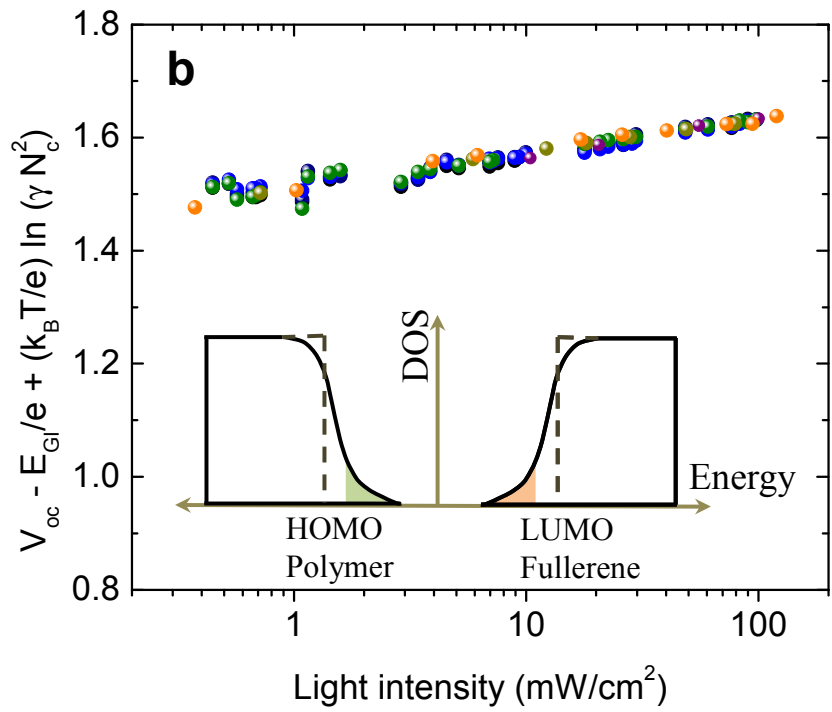

**Figure 3**

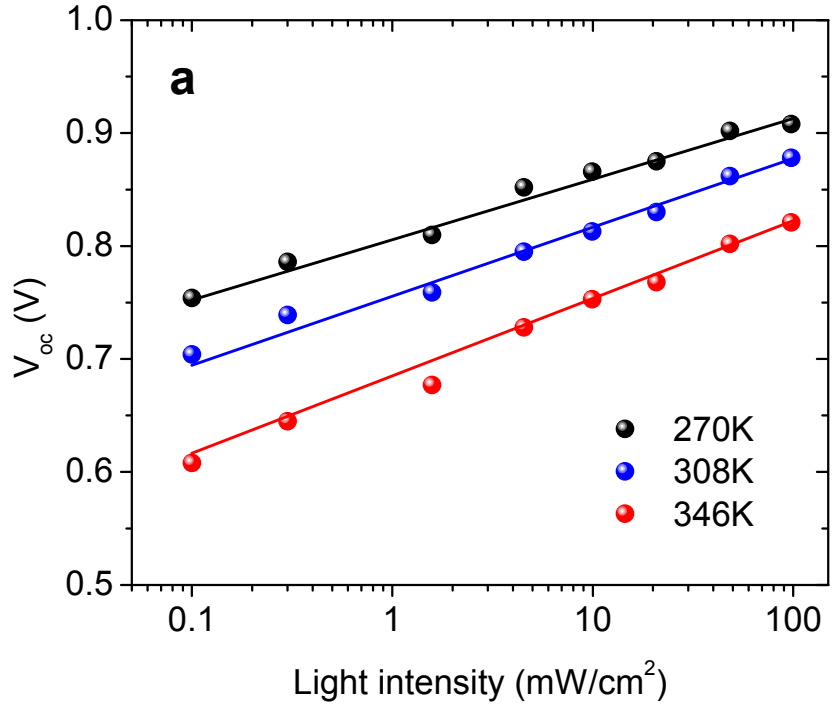
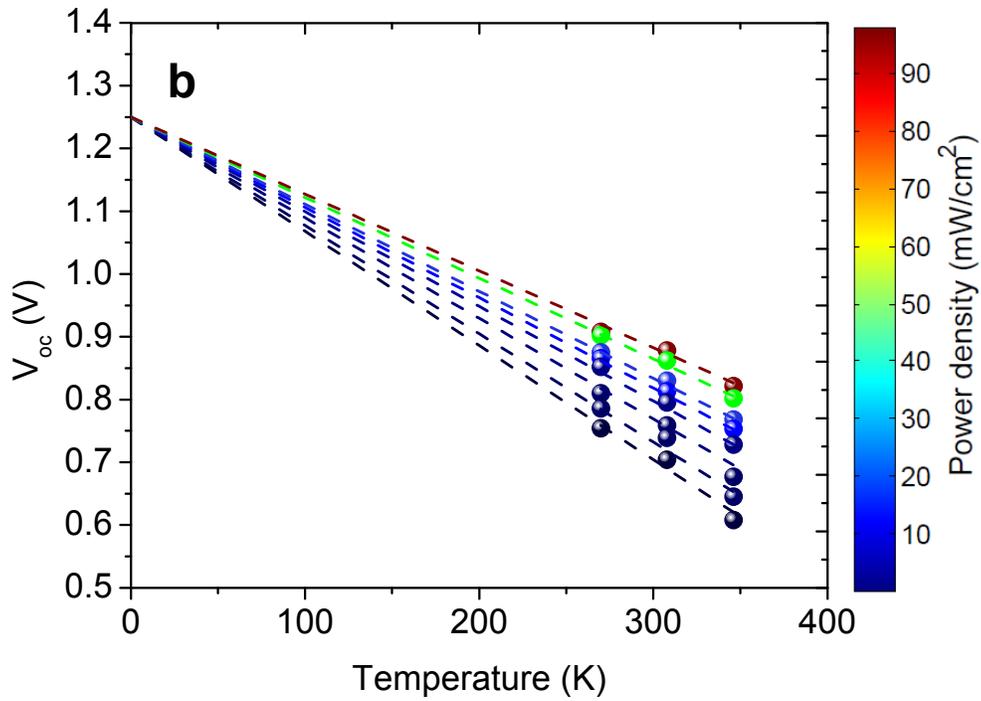

**Figure 4**

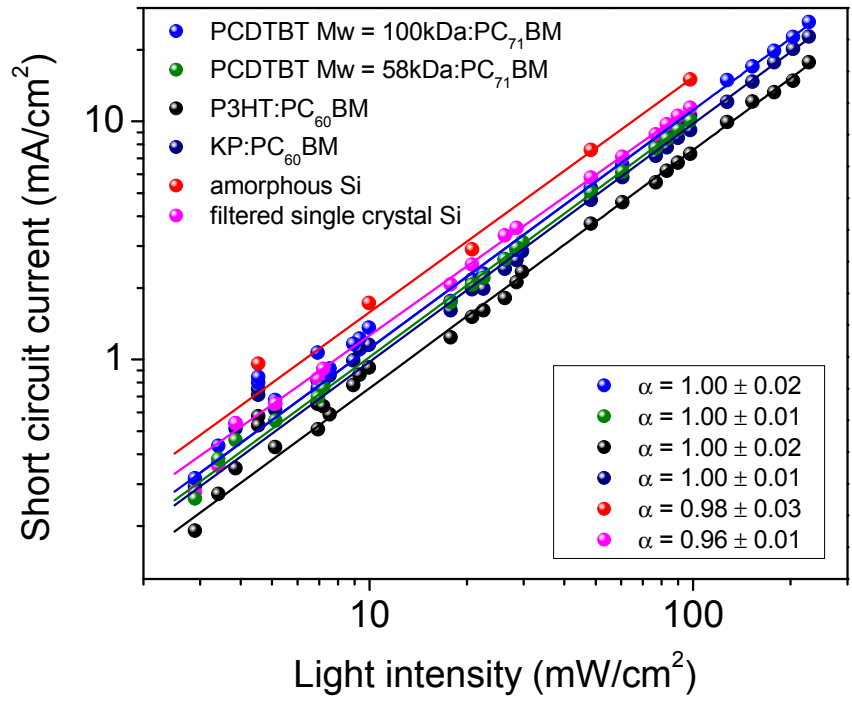

**Figure 5**

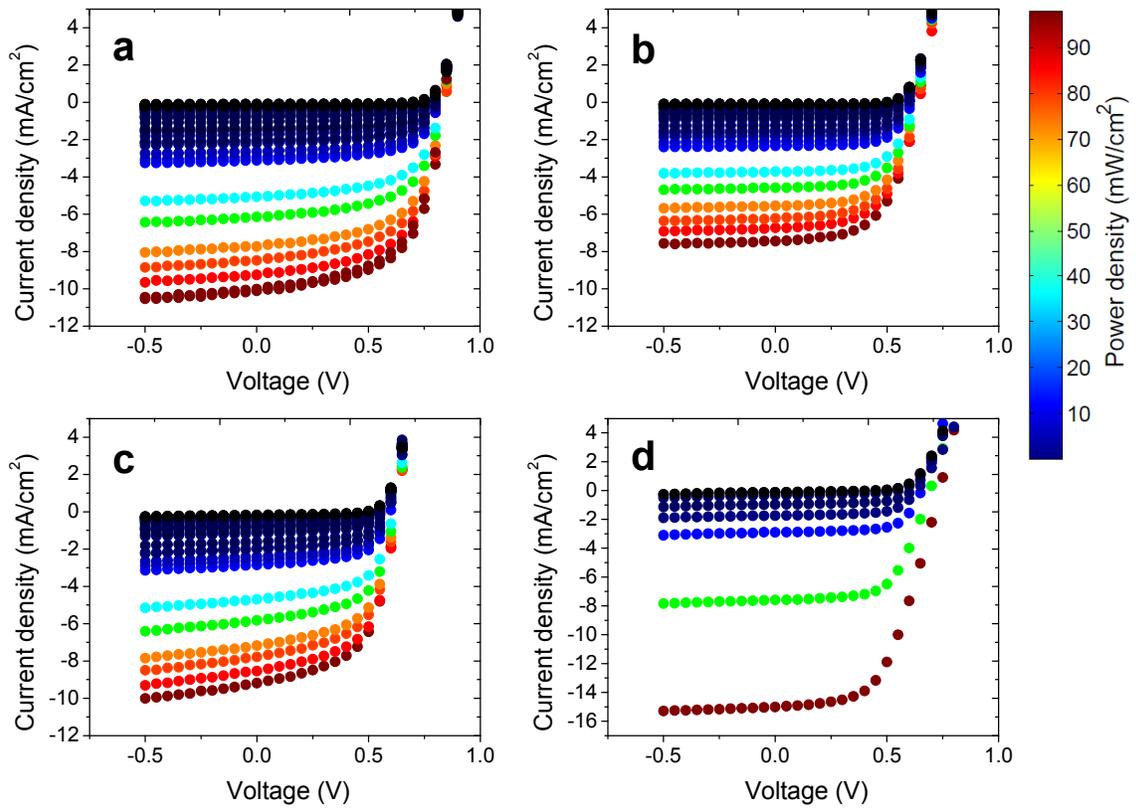

**Figure 6**

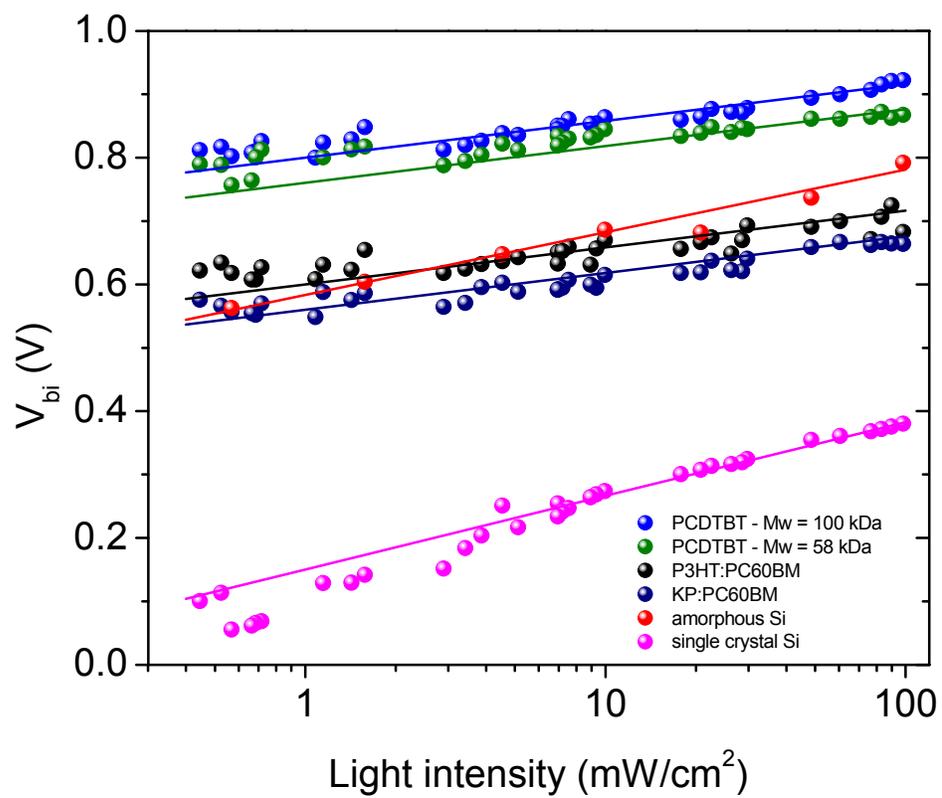

**Figure 7**

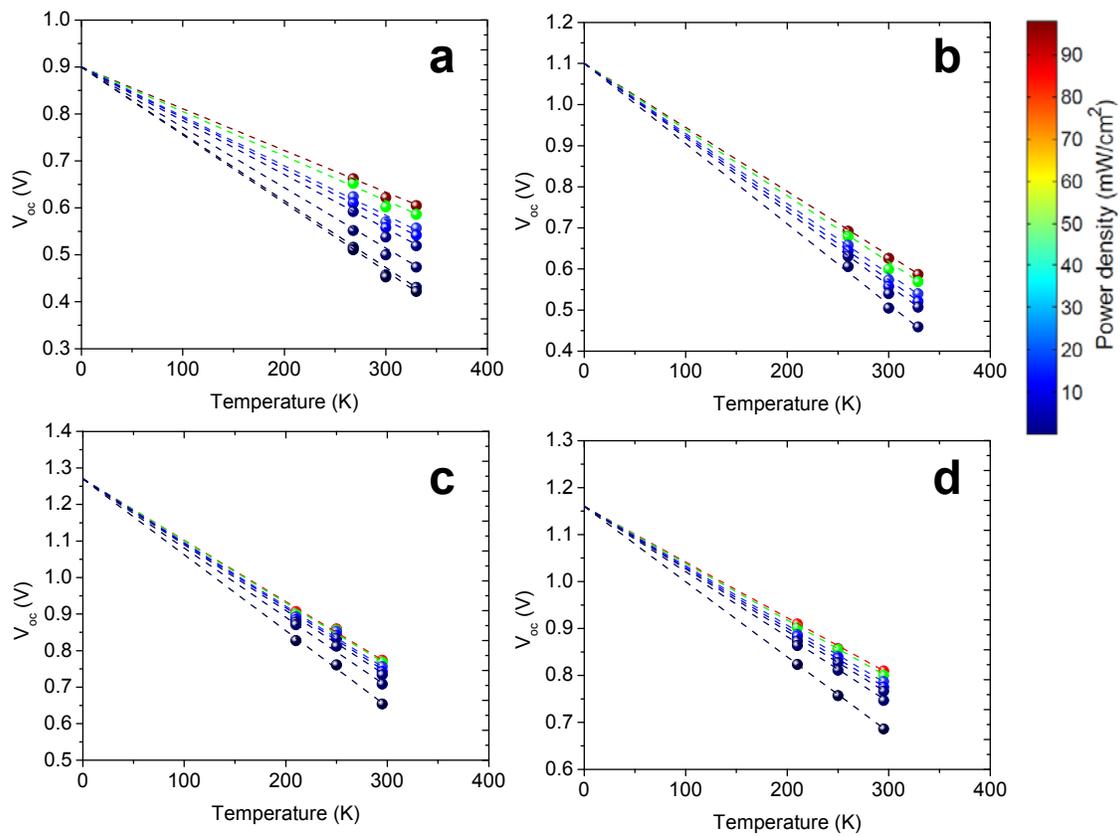

**Figure 8**